\documentclass[a4paper,11pt]{article}
\usepackage{pos}
\usepackage{physics}
\usepackage{cleveref}

\newcommand{\hc}{\mathrm{h.c.}} 
\newcommand{\SO}{\mathrm{SO}}
\newcommand{\SU}{\mathrm{SU}}
\newcommand{\U}{\mathrm{U}}
\newcommand{\Sp}{\mathrm{Sp}}

\title{Colour sextet baryons in composite Higgs models}

\author*{Manuel Kunkel}

\affiliation{Institut für Theoretische Physik und Astronomie, Julius-Maximilians-Universität Würzburg,\\ Emil-Hilb-Weg 22, 97074 Würzburg, Germany}

\emailAdd{manuel.kunkel@uni-wuerzburg.de}

\abstract{Composite Higgs models with underlying fermionic descriptions predict a rich spectrum beyond the Standard Model. Besides the colour triplet fermionic resonances that act as top partners, also exotically coloured baryons can appear. In this contribution we summarise the phenomenology of colour sextet baryons. 
The dominant decay channel proceeds via a colour octet scalar, resulting in pair production signatures with four or six top quarks.
In an alternative mass hierarchy, the sextets produce bottom quarks and a collider-stable colour singlet.
We derive recasting bounds on the sextet mass, which reach up to $2.64$~TeV.}

\FullConference{14th Edition of the Large Hadron Collider Physics (LHCP2026)\\
18-22 May 2026\\
Paris, France\\}

\begin{document}
\maketitle

\section{Introduction}\label{sec:introduction}

Composite Higgs models provide a well-motivated solution to the naturalness problem of the Standard Model (SM) by positing the Higgs as a pseudo Nambu-Goldstone boson (pNGB) \cite{Kaplan:1983fs,Kaplan:1983sm,Dugan:1984hq}. 
A new asymptotically free gauge theory of hyperquarks becomes strongly interacting around the TeV scale, triggering a spontaneous breaking of the hyperquark flavour group $G$ to a subgroup $H$. 
The hyperquarks form baryonic bound states, among them partners of the third generation quarks in order to generate the large top Yukawa coupling via the partial compositeness mechanism \cite{Kaplan:1991dc}.

We consider a class of twelve promising models with hyperquarks in two distinct irreducible representations (irreps) of the hypercolor gauge group \cite{Ferretti:2013kya,Ferretti:2016upr,Belyaev:2016ftv}: $\psi$ charged only under $\SU(2)_L \times \SU(2)_R$, and $\chi$ carrying $\SU(3)_c$ and $\U(1)_X$ charges, where the hypercharge is built from $Y = T^3_R + X$.
The symmetry breaking pattern only depends on the reality of the irrep.
In the colour sector, the the three minimal cases are $\SU(6)/\SO(6)$ for $\chi$ in a real, $\SU(6)/\Sp(6)$ for a pseudoreal and $\SU(3)^2/\SU(3)$ for a complex irrep.

The models predict a rich and testable zoo of states beyond the Standard Model (BSM) including an extended scalar sector \cite{Belyaev:2015hgo, Cacciapaglia:2015eqa, Belyaev:2016ftv, Ferretti:2016upr, Agugliaro:2018vsu, Cacciapaglia:2019bqz, Cornell:2020usb, Cacciapaglia:2020vyf, BuarqueFranzosi:2021kky, Cacciapaglia:2022bax,Flacke:2023eil, Ferretti:2025zsq,Flacke:2025xwl}, fermionic resonances \cite{DeSimone:2012fs,Buchkremer:2013bha,Bizot:2018tds,Cacciapaglia:2019zmj,Xie:2019gya,Benbrik:2019zdp,Wang:2020ips,Cacciapaglia:2021uqh,Corcella:2021mdl,Banerjee:2022izw,Banerjee:2022xmu,Banerjee:2024zvg,Flacke:2026fxb}, and spin-1 resonances \cite{BuarqueFranzosi:2016ooy,Caliri:2024jdk,Cacciapaglia:2024wdn}.
The electroweak condensate $\expval{\psi\psi}$ contains the Higgs field as well as further pNGBs with electric charge up to 2, whereas the colour sector condensate $\expval{\chi\chi}$ gives rise to a neutral octet and a sextet/triplet for a real/pseudoreal irrep. 
The baryons are of the form $\chi\psi\chi$ or $\psi\chi\psi$.
The latter models only contain vector-like quarks (VLQs), but the former contain baryons of exotic colour representations as well, such as sextets.

In this contribution we summarise the phenomenology of the sextet baryons.
We discuss the relevant particle content and the expected spectrum in \cref{sec:spectrum}, then highlight the dominant decay channels in \cref{sec:decays}.
In \cref{sec:bounds} we present the current LHC bounds on colour sextet pair production, then draw conclusions in \cref{sec:conclusions}.
We refer to \cite{Cacciapaglia:2026jlv} for more details.

\section{Typical spectrum}\label{sec:spectrum}

For simplicity we focus on models with $\SU(6)/\SO(6)$ breaking in the colour sector, corresponding to models M1 and M2 in the nomenclature of \cite{Belyaev:2016ftv}, and discuss generalization to other cosets in \cref{sec:conclusions}.
Besides the ubiquitous neutral colour octet pNGB $\pi_8$, there is also a sextet pNGB $\pi_6$ with electric charge $-2/3$.
The electroweak sector forms a $\SU(5)/\SO(5)$ coset, yielding several neutral scalars $S^0$, two charged states $S^\pm$ and one doubly charged scalar $S^{\pm\pm}$.
We expect the coloured and the EW pNGBs to each have a similar mass with a small splitting between different $\SU(3)_c$ and EW representations.
To simplify the following analysis, we neglect the latter splitting and take the pNGBs to have masses $m_\pi$ and $m_S$, respectively.

\begin{figure}
    \centering
    \includegraphics[height=4.5cm]{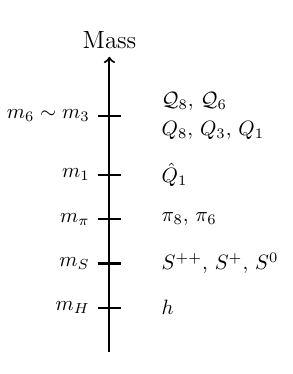} \qquad 
    \includegraphics[height=4.5cm]{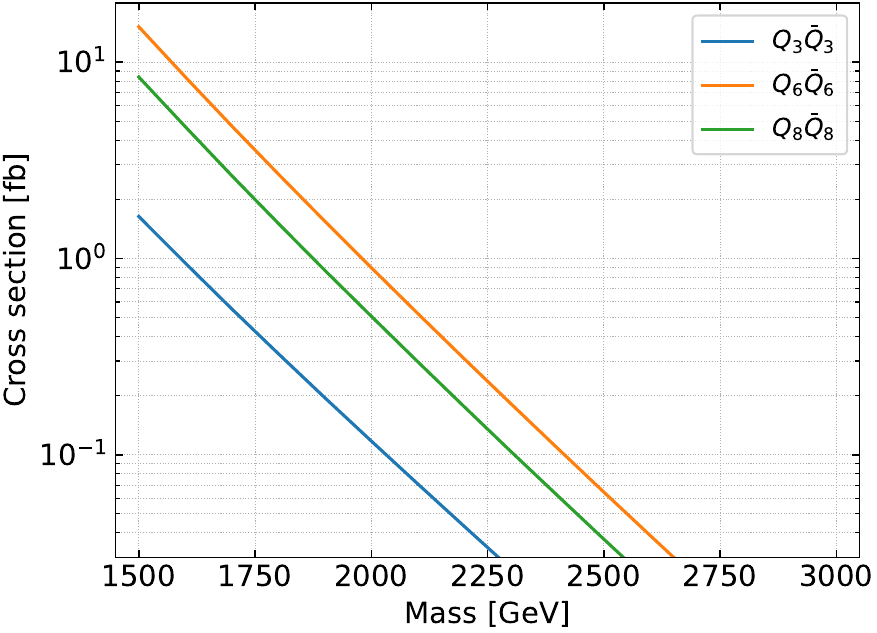}
    \caption{Typical spectrum (left) and cross sections for pair production of baryons (right) at $\sqrt s = 13$~TeV calculated at leading order.}
    \label{fig:spectrum_xs}
\end{figure}

The baryons are of the type $\chi \psi \chi$. The hyperquarks live in $\psi \in (\mathbf 5, \mathbf 1)$ and $\chi \in (\mathbf 1, \mathbf 6)$ of the global flavour group $\SU(5) \times \SU(6)$.
Decomposing to the unbroken subgroup $\SO(5) \times \SO(6)$ yields 
\begin{align}
    \chi \psi \chi \to (\mathbf 5, \mathbf{20}) + (\mathbf 5, \mathbf{15}) + (\mathbf 5, \mathbf{1}) = Q_{20} + Q_{15} + Q_1.
\end{align}
Under $\SO(6) \to \SU(3)_c \times \U(1)_X$, $\mathbf{20} \to \mathbf 8_0 + \mathbf 6_{-2/3} + \mathbf{\bar 6}_{2/3}$ and $ \mathbf{15} \to \mathbf 8_0 + \mathbf 3_{2/3} + \mathbf{\bar 3}_{-2/3}$.
Combined with $\mathbf 5_{\SO(5)} \to (\mathbf 2, \mathbf 2) + (\mathbf 1, \mathbf 1)$ of $\SU(2)_L \times \SU(2)_R$, the colour triplets read
\begin{align}
    (X_{5/3}, X_{2/3}) \in (\mathbf 3, \mathbf 2)_{7/6}, \quad (T,B) \in (\mathbf 3, \mathbf 2)_{1/6}, \quad \tilde T \in (\mathbf 3, \mathbf 1)_{2/3} \quad \text{of } \SU(3)_c \times \SU(2)_L \times \U(1)_Y
\end{align}
and we identify $(T,B)$ and $\tilde T$ as the partners for the left- and right-handed top quarks.
The model also contains a VLQ with charge $5/3$.
Besides the colour triplets, the multiplet $Q_{15}$ further contains octets and singlets, each with electric charge $\pm 1,0$ (Dirac) and one Majorana state.
All states in $Q_{15}$ are expected to have a similar mass $m_3$, apart from corrections from QCD and the top mixing.
Rather than triplets and singlets, the $Q_{20}$ gives rise to sextet baryons 
as well as octets, with masses around a common scale $m_6$.
The sextets form five Dirac states,
\begin{align}
    Q_6^{-5/3}, \quad 3\times Q_6^{-2/3}, \quad Q_6^{1/3}.
\end{align}
Finally, a colour singlet $\hat Q_1$ remains, yielding one charged and one neutral Dirac state as well as one Majorana state.
The model has five mass scales: $m_\pi$ and $m_S$ for the pNGBs and $m_{1,3,6}$ for the baryons.
To evade bounds on pNGB pair production, we take $m_S = 500$~GeV \cite{Cacciapaglia:2022bax} and $m_\pi = 1500$~GeV \cite{Kunkel:2025qld}.
We assume $m_1 < m_6 \sim m_3$, leading to the spectrum shown in \cref{fig:spectrum_xs} (left).
The right panel of \cref{fig:spectrum_xs} shows that for a fixed mass, colour sextets have a larger pair production cross section\footnote{For colour sextets, no NLO UFO \cite{Degrande:2011ua, Darme:2023jdn} model implementation is available. We therefore only show leading order cross sections calculated with \texttt{MadGraph5\_aMC@NLO} \cite{Alwall:2014hca} using the \texttt{NNPDF2.3} set of parton densities \cite{Ball:2012cx,Buckley:2014ana} and fixing the renormalisation and factorisation scales to $\mu_R = \mu_F = m_Q$.} than octets, let alone triplets.
The sextet baryons can therefore still be produced with appreciable cross sections into the multi-TeV range, making them an interesting object of study.

\section{Decay channels}\label{sec:decays}

The baryon multiplets are connected through derivative couplings
\begin{align}
    \mathcal L_\mathrm{int} \supset \frac{c_{15}}{f} \bar Q_{20} \, \bar \sigma^\mu \partial_\mu \pi \, Q_{15} + \frac{c_{1}}{f} \bar Q_{20} \, \bar \sigma^\mu \partial_\mu\pi \, \hat Q_{1} + \hc ,
\end{align}
opening a multitude of decay channels: $Q_6 \to \pi_6 Q_8, \, \pi_8 Q_3^c, \, \pi_6 Q_1, \, \pi_6 \hat Q_1$.
Unless we take $m_6 \gg m_3$, most of these decays will be off-shell.
The exception are decays into the VLQs which mix with the third generation quarks.
Thus the dominant decays for the sextet baryons with charge $-2/3$ and $1/3$ are
\begin{align}\label{eq:q6dec}
    Q_6^{-2/3} \to \bar t \pi_8, \qquad Q_6^{1/3} \to \bar b \pi_8.
\end{align}
While the $X_{5/3}$ does not mix with quarks, its decay products $X_{5/3} \to tW^+, \, tS^+ ,\, bS^{++}$
are relatively light, making the three body decays $Q_6^{-5/3} \to \bar X_{5/3}^* \pi_8 $ the kinematically favourable decay channels of the $Q_6^{-5/3}$.
Motivated by partial compositeness, we take the pNGBs to decay to third generation quarks: $ S^0 \to t\bar t$, $S^+ \to t\bar b$, $S^{++} \to W^+ S^* \to W^+ t \bar b$ for the uncoloured states and $\pi_8 \to t\bar t$, $\pi_6 \to bb$ for the coloured pNGBs, where the $\pi_6$ decay violates baryon number.
All in all, we therefore have
\begin{align}
    Q_6^{-2/3} \to \bar t t \bar t, \quad Q_6^{1/3} \to \bar b t\bar t, \quad Q_6^{-5/3} \to \bar tW^+ t\bar t, \, \bar t t\bar b t\bar t,\, \bar b W^+ t\bar b t\bar t.
\end{align}

An alternative scenario to the decays above can occur when the $\hat Q_1$ is significantly lighter than the coloured baryons, such that the decay $Q_6 \to \pi_6 \hat Q_1$ is on-shell and competitive with \cref{eq:q6dec}.
This would require a very small $m_1$, which is not generically expected.
However, we also cover this case for completeness.
The lightest state $\hat q_1^0$ of the $\hat Q_1$ does not have a decay channel and can serve a dark matter candidate. 
The heavier components of $\hat Q_1$ decay into $\hat q_1^0$ and very soft particles that cannot be resolved by the detector, so that the whole $\hat Q_1$ only leaves a signature of missing transverse energy (MET), and we have
\begin{align}
    Q_6 \to bb + \mathrm{MET}.
\end{align}

\section{LHC bounds}\label{sec:bounds}

At hadron colliders, the sextet baryons can be QCD pair produced, leading to the signatures
\begin{align}
    Q_6 \bar Q_6 \to 6t \, / \, 2b 4t \, / \, 6t 2W \qquad \text{or} \qquad Q_6 \bar Q_6 \to 4b+\mathrm{MET}
\end{align}
where we only consider the simplest decay $Q_6^{-5/3} \to \bar t W^-\pi_8$.
We generate hard scattering events at leading order with \texttt{MadGraph5} \cite{Alwall:2014hca} and produce a showered and hadronised \texttt{HepMC} \cite{Dobbs:2001ck} event file with \texttt{Pythia8} \cite{Sjostrand:2014zea}.
Due to technical limitations with colour sextets, we use a surrogate model \cite{Banerjee:2022xmu} with triplet fermions.
We pass the events to \texttt{MadAnalysis5} \cite{Conte:2012fm, Conte:2014zja, Dumont:2014tja, Conte:2018vmg, Araz:2025bww}, which performs detector simulation with \texttt{Delphes 3} \cite{deFavereau:2013fsa} or the \texttt{SFS} framework \cite{Araz:2020lnp, Araz:2021akd}.
The events are run through the cuts of recast analyses, and from the remaining events an exclusion value is calculated with the CL$_s$ method \cite{Read:2002hq,Heinrich:2021gyp,Araz:2023bwx}.
We run the events against all implemented analyses and find \texttt{ATLAS-SUSY-2018-31} \cite{ATLAS:2019gdh, IHALED_2020} and \texttt{ATLAS-SUSY-2018-17} \cite{ATLAS:2020xgt, I2CZWU_2021} to give the dominant bounds for the decay via $\pi_8$, while \cite{ATLAS:2019gdh} and \texttt{CMS-SUS-19-006} \cite{CMS:2019zmd, 4DEJQM_2020} dominate for the $4b+\mathrm{MET}$ signature.

\begin{figure}
    \centering
    \includegraphics[width=0.43\linewidth]{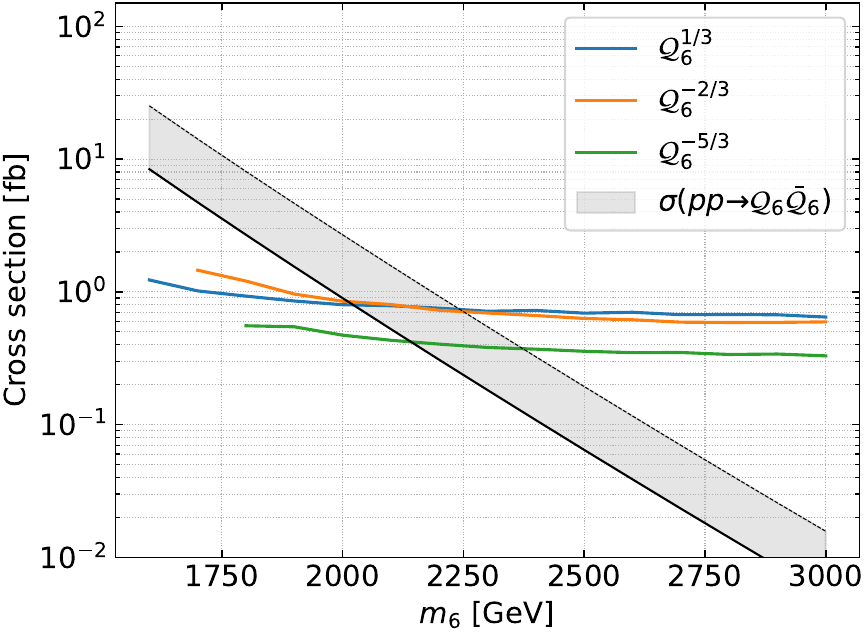} \quad 
    \includegraphics[width=0.43\linewidth]{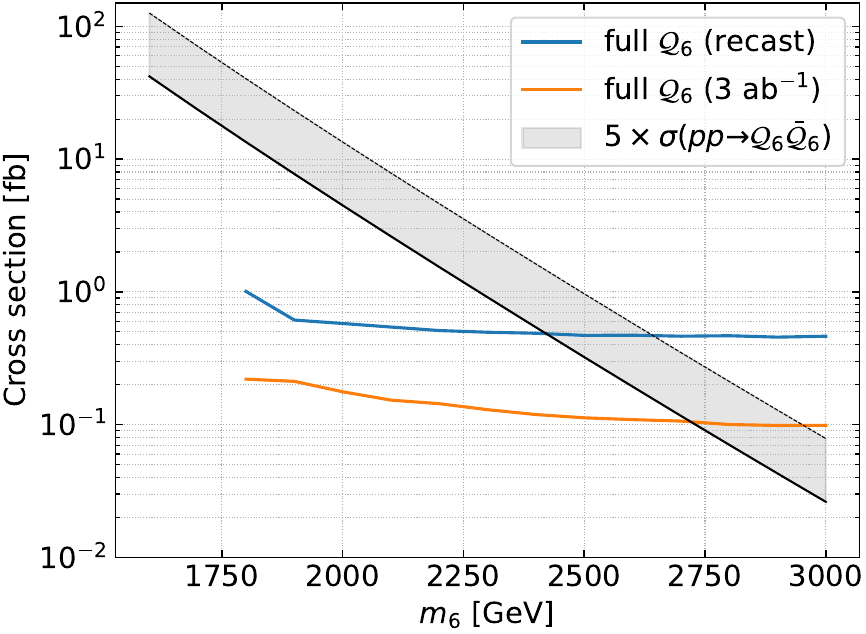}
    \caption{Recasting bounds on pair production of sextet baryons for the $Q_6 \to \bar Q_3 \pi_8$ decay.}
    \label{fig:bounds_1dim}
\end{figure}

\begin{figure}
    \centering
    \includegraphics[width=0.43\linewidth]{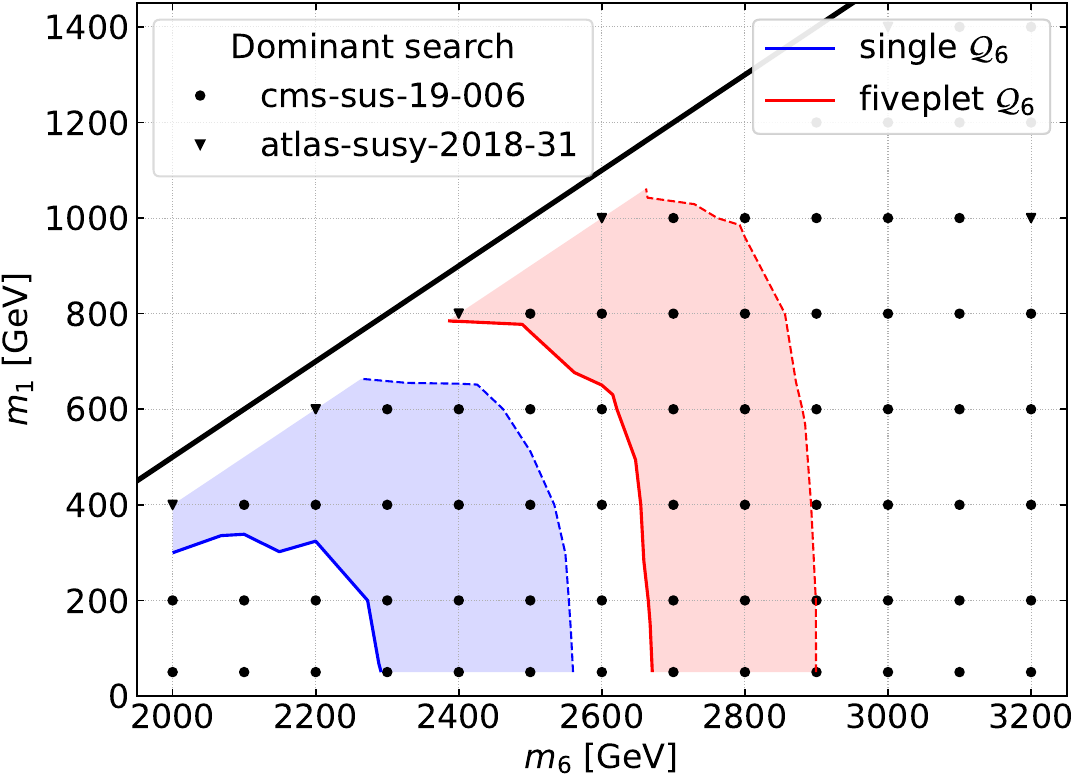} \quad 
    \includegraphics[width=0.43\linewidth]{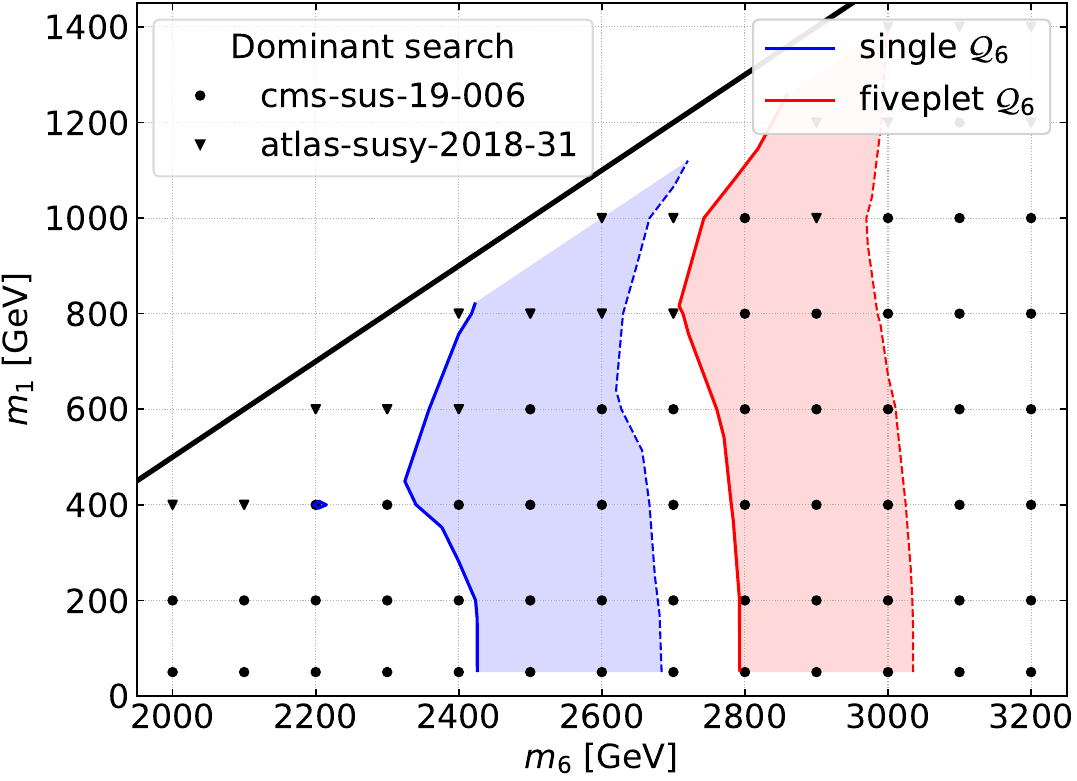}
    \caption{Recasting bounds on pair production of sextet baryons for the $Q_6 \to bb+\mathrm{MET}$ decay.}
    \label{fig:bounds_met}
\end{figure}

In \cref{fig:bounds_1dim} we show the bounds on the $Q_6 \to \bar Q_3 \pi_8$ decay for the individual states (left) and for the whole multiplet (right). 
The coloured lines indicate the observed upper limits on the cross section.
The grey band shows the production cross section between the leading order result (black line) and a flat $K$-factor of $K=3$, taken to signify our ignorance of the NLO cross section.
The full multiplet is excluded up to $2.64$~TeV from recasts using 139~fb$^{-1}$ of data. 
Extrapolating to 3~fb$^{-1}$, this can be extended to almost 3~TeV.
If we allow for very small $m_1$, the $Q_6 \to bb + \mathrm{MET}$ decay becomes important.
Bounds on this channel in the $m_6$-$m_1$-plane are shown in \cref{fig:bounds_met}, excluding sextet masses up to 2.9~TeV for 139~fb$^{-1}$ (left) and up to $3.05$~TeV when projected to 3~ab$^{-1}$ (right).

\section{Conclusions}\label{sec:conclusions}

We showed that composite Higgs models can produce colour sextet baryons, which typically decay into multi-top final states. 
We determined recast bounds from Run-2 searches, setting a lower limit of 2.64~TeV on the sextet mass, and have extrapolated the bounds to the HL-LHC.
While we only discussed the $\SU(6)/\SO(6)$ cosets, the results also apply to $\SU(6)/\Sp(6)$ and $\SU(3)^2/\SU(3)$, except the $Q_6 \to bb+\mathrm{MET}$ channel is absent since these coset don't contain a sextet pNGB.

\acknowledgments
I thank Giacomo Cacciapaglia, Rosy Caliri, Aldo Deandrea, Benjamin Fuks, Mark Goodsell, Jan Hadlik and Werner Porod for collaboration on this project.
I am supported by the DFG research training group GRK 2994.
This work was supported by the German Academic Exchange Service (DAAD), PROCOPE project nr. 57755908.

\bibliographystyle{utphys}
\bibliography{literature}

\end{document}